\documentclass{article} 
\usepackage{authblk}

\usepackage[utf8]{inputenc}
\usepackage[margin=1in]{geometry}
\usepackage{amsmath}
\usepackage{amssymb}
\usepackage{hyperref}
\usepackage{physics}
\usepackage{tikz}
\usepackage{bm}

\begin{document}

\title{Numerical simulation of non-central collisions of spherical magnets}
\author{Sean P.~Bartz\footnote{sean.bartz@indstate.edu} \and Jacob Shaw}
%\email{sean.bartz@indstate.edu}
%\author{Jacob Shaw}
%\affiliation{Dept.~of Chemistry and Physics, Indiana State University, Terre Haute, IN 47809}
\date{ 
Dept.~of Chemistry and Physics, Indiana State University, Terre Haute, IN 47809\\
\today}

\maketitle

\begin{abstract}
We present a computational model of non-central collisions  of two spherical neodymium-iron-boron magnets, suggested as a demonstration of angular momentum conservation.  Our program uses an attractive dipole-dipole force and a repulsive contact force to solve the Newtonian equations of motion for the magnets. We confirm the conservation of angular momentum and study the changes in energy throughout the interaction.  Using the exact expression for the dipole-dipole force, including non-central terms, we correctly model the final rotational frequencies, which is not possible with a simple power-law approximation.
\end{abstract}

\section{Introduction}
An experimental paper by Lind\`{e}n et al  \cite{Lindn2018} showed that two spherical NdFeB magnets projected toward each other with a nonzero impact parameter end up revolving around each other with a large angular velocity.  
In this experiment, the magnets are rolled down ramps toward each other, and collide in mid-air. The impact velocity is measured using high-speed cameras, and the angular velocity  of the final state is measured by a pick-up coil connected to an oscilloscope. The measured angular velocities match the values predicted by   angular momentum conservation. 
The authors also suggest a follow-up analysis of the energy of the system.

 In this paper, we use theoretical and computational techniques suitable for undergraduate students to explore this interaction.
Numerical techniques in the undergraduate curriculum typically focus on the Newtonian approach of updating an object's velocity and position from the sum of forces acting on it \cite{Weber2020}. Computation allows students to extend this paradigm to physical situations where closed form solution is  difficult or intractable. 

Conservation laws offer an alternative to the Newtonian approach that is particularly useful in comparing initial and final states. The problems considered in this paper are amenable to solution by considering angular momentum and energy conservation, but the computational solution of the magnets' trajectories aids in student understanding, particularly through the animation of the simulation.

We begin by approximating the dipole-dipole force as a power law, and using a damped spring as the contact force between the spheres. The spheres collide and revolve as expected, but the final angular velocity of the ``barbell'' shape does not match the prediction from angular momentum considerations. 
This shortcoming is corrected by incorporating the complete force and torque expressions for the dipole-dipole interaction, which includes non-central terms.

% Information regarding student research is discussed in this paper to highlight not only the discoveries found regarding energy of revolving neodymium spheres, but also the learning process to conducting research in a theoretical approach.
% The spring force, dissipative force, and dipole force the two magnetic spheres were made dimensionless. 
Non-dimensionalization is an important technique for students to learn for numerical analysis. This technique helps to identify appropriate spatial and time scales for the simulation, while also reducing the number of input parameters. 
We reduce the physical parameters of the system to a dimensionless damping ratio and the radius of the magnets, expressed in terms of a characteristic distance. 
The contact force acts solely in the radial direction, so it does not affect the angular momentum. 
The choice of damping ratio determines how many times the magnets bounce off of each other, but we are only concerned with the final rotational state once the bouncing has ceased.
Thus, the final state does not depend on the parameters that characterize the contact force.
We  focus on the effect of the impact parameter and the initial kinetic energy on the final angular velocity and kinetic energy of the system.

% The impact of different initial conditions of the spheres were able to me witnessed and how these alterations impacted the various energies. We changed the impact parameter as well as the dissipative coefficient while ensuring that the initial energy remained constant. 

\section{Dipole-dipole interaction}
It has been shown \cite{Edwards2017a} that the interaction of two spherical magnets of uniform magnetization is equivalent to the interaction of pure magnetic dipoles. 
The potential energy of a magnetic dipole interacting with a magnetic field $\vb{B}$ is
\begin{equation}
    U=-    \bm{\mu} \cdot \vb{B}, \label{eq:Udef}
\end{equation}
where $   \bm{\mu}$ is the magnetic dipole moment.  
\begin{figure}
    \centering
\begin{tikzpicture}
% \coordinate (origin) at (0,0)
% \coordinate (c1) at (2,3)
% \coordinate (c2) at (6,4)

\draw[](2,3) circle (1.1);
\draw[](6,4) circle (1.1);
\draw[->,>=stealth,thick](2,3) -- node[sloped, anchor=south]{$\vb{r}$} (6,4) ;
\draw[->,>=stealth,thick](4,0) --node[ anchor=east]{$\vb{r}_1$} (2,3);
\draw[->,>=stealth, thick](4,0) --node[ anchor=west]{$\vb{r}_2$} (6,4);
\draw[->,>=stealth, thick](4,0) --node[ anchor=west]{$\vb{r}_\mathrm{com}$} (4,3.5);
\draw[blue,dashed] (2,3) --(3.6,3);
\draw[blue, dashed] (6,4) --(7.1,4);
\draw[->,>=stealth] (2,3) -- (1.5,4.5) node[anchor=west]{$\bm{\mu}_1$};
\draw[blue] (2.5,3) arc (0:107:0.5) node[midway, anchor=south west]{$\phi_1$};
\draw[->,>=stealth] (6,4) -- (7,5) node[anchor=west]{$\bm{\mu}_2$};
\draw[blue] (6.5,4) arc (0:45:0.5)  node[anchor=west]{$\phi_2$};
\draw[red] (3.3,3) arc (0:15:1.3) ;
\draw[red] (3.35,3) arc (0:15:1.35) node[anchor=west, midway]{$\theta$} ;
\end{tikzpicture}
    \caption{Schematic illustrating the coordinate definitions, including the orientation of the magnetic dipole moments $\bm{\mu}_1, \, \bm{\mu}_2$. We restrict the initial conditions such that the magnetic moments lie in the collision plane of the spheres.}
    \label{fig:coordsDef}
\end{figure}
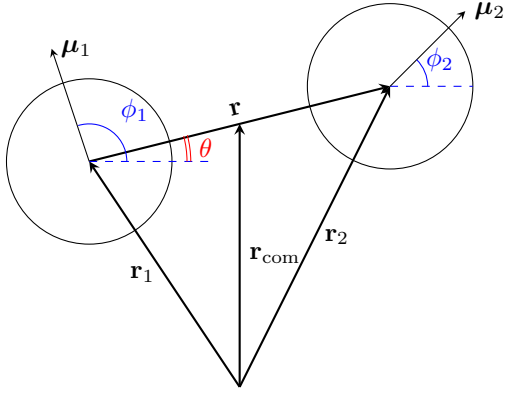
Let us find the potential energy of a dipole $\bm{\mu}_1$ interacting with the magnetic field of dipole $\bm{\mu}_2$. The magnetic field at the location of dipole $\bm{\mu}_2$ due to dipole $\bm{\mu}_1$ is  \cite{Griffiths_EM}
\begin{equation}
\vb{B}(\vb{r}_{2})=\frac{\mu_0}{4\pi r ^3} [  3(    \bm{\mu}_1 \cdot \vu{r} )\vu{r} -    \bm{\mu}_1  ], \label{eq:Bdef}
\end{equation}
where $\vb{r} =\vb{r}_2-\vb{r}_1$ is the relative position of the two magnets, as shown in Figure \ref{fig:coordsDef}, and $\vu{r}$ is the unit vector
\begin{equation}
    \vu{r}=\frac{\vb{r}_2-\vb{r}_1}{\left| \vb{r}_2-\vb{r}_1 \right|}.
\end{equation}
The potential energy calculated using (\ref{eq:Udef}) and (\ref{eq:Bdef}) is
\begin{equation}
  U_{12}=  -\frac{\mu_0}{4\pi r ^3} \left[  3(\bm{\mu}_1 \cdot \vu{r} )(\bm{\mu}_2 \cdot \vu{r}) -    \bm{\mu}_1 \cdot \bm{\mu}_2 \right]. 
  \label{eq:PE}
\end{equation}
Expanding the dot products in terms of the angles defined in Figure \ref{fig:coordsDef}, this expression becomes
\begin{equation}
    U_{12}=-\frac{\mu_0\mu_1\mu_2}{8\pi r ^3}\left[ 3\cos(2\theta -\phi_1-\phi_2) +\cos(\phi_1-\phi_2)    \right]. \label{eq:PEangles}
\end{equation}

The force on the dipole is found from the gradient of the potential energy \cite{Greene1971, Boyer1988}
\begin{equation}
    \vb{F}=\grad(    \bm{\mu}\cdot \vb{B}).
\end{equation}
The generic expression for the force on $\bm{\mu}_2$ is \cite{Yung1998}
% \begin{equation}
    % \vb{F}_1= \frac{3\mu_0}{4\pi r^4} \left ( (\vu{r}\times     \bm{\mu}_1)\times    \bm{\mu}_2 + (\vu{r}\times     \bm{\mu}_2)\times    \bm{\mu}_1 -2(     \bm{\mu}_1\cdot     \bm{\mu}_2)\vu{r} -5(\vu{r}\times     \bm{\mu}_1)\cdot(\vu{r} \times     \bm{\mu}_2)\vu{r} \right ). \label{eq:FullForceCross}
%  \end{equation}
%  which is equivalent to 
\begin{equation}
     \vb{F}_1= \frac{3\mu_0}{4\pi r^4} \left[(    \bm{\mu}_1 \cdot     \bm{\mu}_2)\vu{r} 
     -5(    \bm{\mu}_1 \cdot \vu{r})(    \bm{\mu}_2 \cdot \vu{r}) \vu{r}+
     (    \bm{\mu}_1 \cdot \vu{r})    \bm{\mu}_2 + (    \bm{\mu}_2 \cdot \vu{r})    \bm{\mu}_1  \right]. \label{eq:FullForceDot}
\end{equation}

\subsection{Torque}
The angular momentum of the system arises from two contributions --  an orbital contribution $\vb{L} $ and the angular momentum from the spinning of the spheres $ \vb{S}$.
% \begin{equation}
%      \vb{L}=\vb{L}_\mathrm{rev}+\vb{L}_\mathrm{rot}.
% \end{equation}
We calculate $\vb{L}$ using the definition of angular momentum for a point mass $m$ around a given point $\vb{r}_\mathrm{com}$
\begin{equation}
     \vb{L} = m \Dot{\vb{r}}_1 \cross (\vb{r}_1-\vb{r}_\mathrm{com}) +m \Dot{\vb{r}}_2 \cross (\vb{r}_2-\vb{r}_\mathrm{com}).
\end{equation}
In this paper, we consider only cases where the center of mass is stationary, so we define $\vb{r}_\mathrm{com}$ as the origin for convenience. By definition, $\vb{L}$ is perpendicular to the collision plane.

The rotational contribution arises from considering the spheres as rigid objects. We restrict our analysis to initial conditions where the dipole moments lie in the collision plane, and the magnets are not initially spinning. Thus, the spin angular momentum is is perpendicular to the collision plane, and 
\begin{equation}
    \vb{S} = I_1 \Dot{\phi}_1 +I_2 \Dot{\phi}_2,
\end{equation}
where the angles $\phi_i$ are defined as shown in Figure \ref{fig:coordsDef} and the dot represents a time derivative. The moments of inertia are $I_i=\beta m R^2$, where the masses $m$ and radii of the spheres $R$ are assumed identical. For solid spheres, $\beta=2/5$.

It is important to note that the experimental setup \cite{Lindn2018} produces an additional spin angular momentum in the collision plane due to the spheres rolling down ramps before the collision. This rolling motion is not included in our simulation. However, as argued in that reference, these rolling spins are anti- aligned, so do not contribute to the overall angular momentum.

Following \cite{Edwards2017a}, we note that the torque on the magnets arises from two separate contributions. The first is the interaction between the dipole and the magnetic field
\begin{eqnarray}
        \boldsymbol\tau_{A1} &=&     \bm{\mu}_1\times \vb{B}_2 \nonumber \\
        &=& \frac{\mu_0}{4\pi r^3} \left[3(\bm{\mu}_2\cdot\vu{r})(    \bm{\mu}_1 \times \vu{r}) -    \bm{\mu}_1 \times     \bm{\mu}_2 \right]. \label{eq:torqueA}
\end{eqnarray}
This torque affects the rotational angular momentum,
\begin{equation}
    \bm{\tau}_{A1}=\Dot{\vb{S}}=I_1\Ddot{\phi}_1.
\end{equation}

There is also a torque from the non-central terms of the force from dipole 2. Using (\ref{eq:FullForceDot}), we calculate
\begin{eqnarray}
    \boldsymbol\tau_{B1} &=& \vb{r}\times \vb{F}_1 \nonumber \\
    &=& \frac{\mu_0}{4\pi r^3} \left[ (    \bm{\mu}_1 \cdot \vu{r})(\vu{r} \times     \bm{\mu}_2) + (    \bm{\mu}_2 \cdot \vu{r})(\vu{r}\times     \bm{\mu}_1) \right]. \label{eq:torqueB}
\end{eqnarray}
This torque affects the orbital motion of the spheres.

In general, $\boldsymbol\tau_{A1}+\boldsymbol\tau_{A2}\neq 0$ and $\boldsymbol\tau_{B1}+\boldsymbol\tau_{B2}\neq 0$, but the sum of all these torques is zero.  Thus,  $\vb{L}$ and $\vb{S}$ both change, but the total angular momentum is conserved.

\section{Angular momentum and energy conservation}

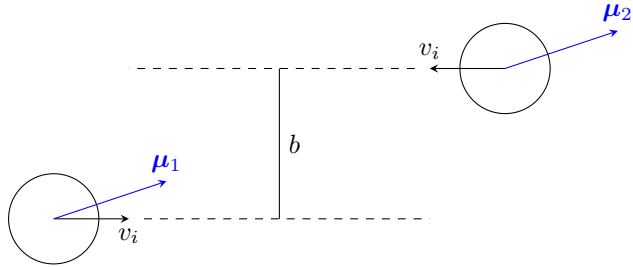
\begin{figure}
    \centering
    \begin{tikzpicture}
    \draw[](0,2) circle (.6);
    \draw[](6,4) circle (.6);
    \draw[->,>=stealth] (0,2) -- (1,2) node[anchor=north]{${v}_i$};
    \draw[->,>=stealth] (6,4) -- (5,4) node[anchor=south]{${v}_i$};
    \draw[] (3,2) -- node[anchor=west]{$b$}(3,4);
    \draw[dashed] (1.2,2) -- (5,2);
    \draw[dashed] (4.8,4) -- (1,4);
    \draw[blue, ->, >=stealth] (0,2)--(1.5,2.5) node[anchor=south]{$\bm{\mu}_1$};
    \draw[blue, ->, >=stealth] (6,4)--(7.5,4.5) node[anchor=south]{$\bm{\mu}_2$};
    \end{tikzpicture}
    \caption{Schematic showing the initial conditions for the magnets. The spheres have identical masses, radii, and initial velocity. The impact parameter $b$ is defined as shown. In this analysis, we examine initial conditions where magnetic moments $\bm{\mu}_1, \, \bm{\mu}_2$ are aligned with the relative position vector $\vu{r}$, defined in Figure \ref{fig:coordsDef}. \label{fig:initialConditions}}
    \label{fig:my_label}
\end{figure}

 Two identical spheres of mass $m$ and radius $R$ are projected toward each other with an initial velocity $v_i $ and impact parameter $b$, as shown in Figure \ref{fig:initialConditions}.
 The initial separation is large enough that the dipole-dipole interaction is negligible. 
 In the final state, the spheres are assumed to form a rigid ``barbell" shape that revolves with an angular velocity $\omega_f$.

The initial angular momentum of the system about the center of mass is $\vb{L}_i=m v_i b$.  The revolving barbell has angular momentum $\vb{L}_f = I \omega_f,$ where $I=\beta m R^2$ is the moment of inertia for the system, and $\beta=14/5$. 
 The angular velocity of this barbell shape is \cite{Lindn2018}
% \begin{eqnarray}
%     \vb{L}_f &=& \vb{L}_i \\
%     I \omega_f &=& M v_i b \\
%     \omega_f &=& \frac{M v_i b}{\beta M R^2} \\
%     \omega_f &=& \frac{ v_i b}{\beta  R^2}
% \end{eqnarray}
\begin{equation}
    \omega_f = \frac{ v_i b}{\beta  R^2}. \label{eq:omegaf}
\end{equation}

With the angular velocity determined, we can calculate the ratio of the final kinetic energy $K_f=\frac{1}{2} I \omega_f^2$ to the initial kinetic energy of the system $K_i=m v_i^2$. 
\begin{equation}
    \frac{K_f}{K_i} = \frac{1}{2\beta}\left( \frac{b}{R} \right) ^2 \label{eq:KEratio}
\end{equation}
Thus, the final kinetic energy exceeds the initial when $b/R>1/\sqrt{2\beta}$. 

The relationships (\ref{eq:omegaf}), (\ref{eq:KEratio}) do not depend on the details of the attractive force between the spheres. However, these apply only if the spheres end up stuck together. 
Whether the spheres make contact depends on the characteristics of the attractive force, as well as the initial conditions of the motion. 
The details of the contact force determine whether spheres that make contact stick together rather than bouncing apart. 
Theoretical and numerical investigation of these conditions is reserved for future work.
In this paper, we restrict our investigation to situations that produce a final rotating barbell.

% We consider the effective potential for the dipole-dipole interaction with 
% \begin{equation}
%     U_{eff}=-\frac{\mu_0 m_1 m_2}{2\pi r ^3 }+ \frac{M\ell^2}{2r ^2}
% \end{equation}

% If we have
% \begin{equation}
%     U_{eff}=-\frac{\alpha}{r^3}+\frac{\mu \ell^2}{2r^2},
% \end{equation}
% where $\mu$ is the reduced mass, then the maximum velocity a particle can have and be captured is
% \begin{equation}
%     v=\frac{\sqrt{\alpha}3^{3/4}}{\sqrt{\mu} b^{3/2}}
% \end{equation}
% However, we find a slope that is $\sim$1.63 times the expected value.
\section{Numerical solution \label{fullforce}}
In addition to the dipole-dipole interaction, we model the contact force between the two spheres as a spring with a linear damping term. 
This force law is based on the Kelvin-Voigt model, which describes materials as elastic on long time scales, but rapid deformation results in an additional resistance 
\cite{meyers_chawla_2008}.
This simple force law has been successfully applied  to inelastic collisions of spheres \cite{Nagurka, Muller2011} . For oblique collisions of smooth particles, both the elastic and dissipative forces point along $\vb{r} $, in the normal direction \cite{Becker2008}.

\begin{figure}[htb]
    \centering
    \includegraphics[width=0.8\textwidth ]{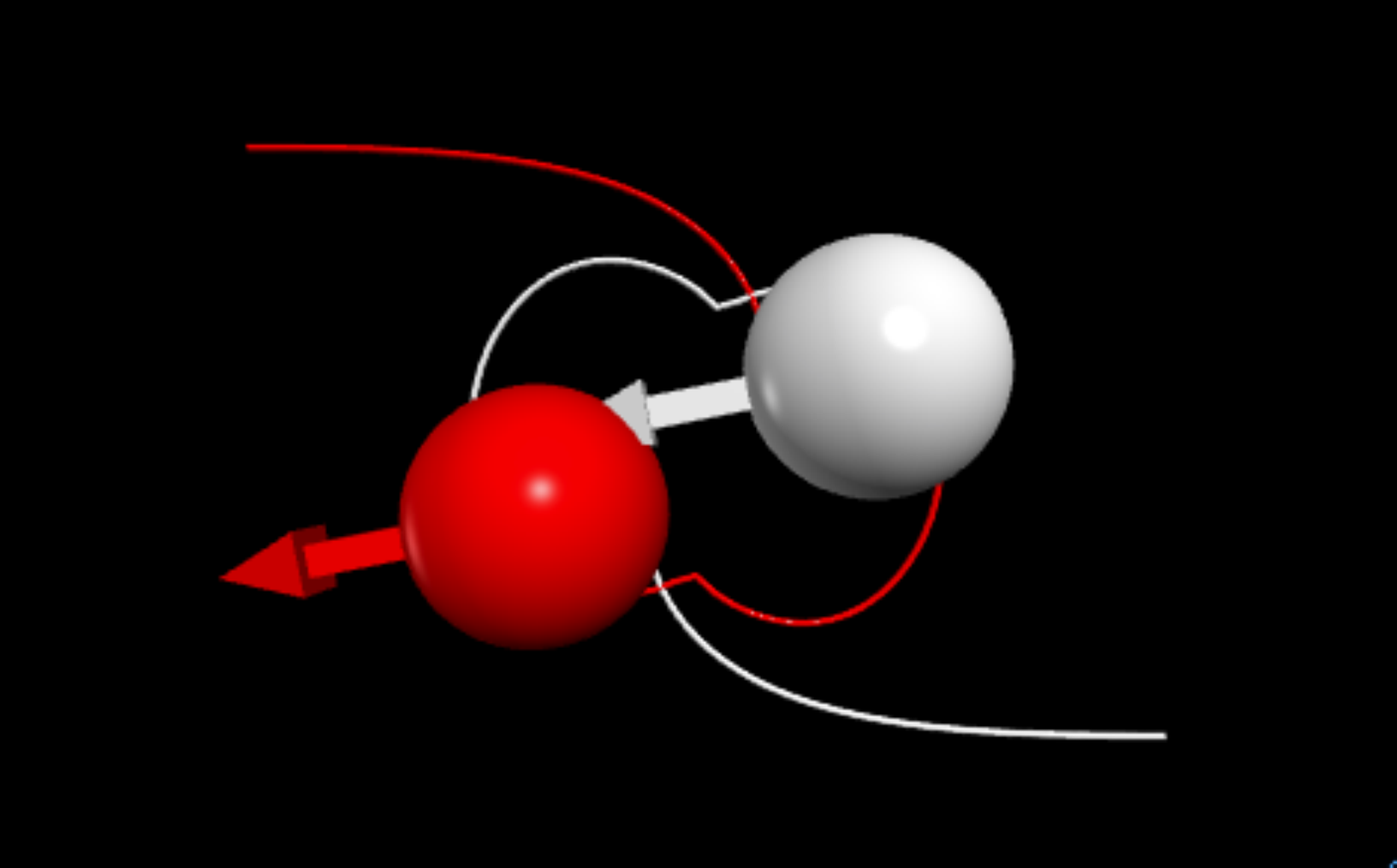}
    \caption{A typical animation of the colliding magnets produced by VPython, showing the paths of the spheres. The arrows indicate the orientations of the magnetic moments $\vb{\mu}_i$.}
    \label{fig:Sphere_Animation}
\end{figure}

The damped spring force is defined 
\begin{equation}
    \vb{F}_s = \begin{cases}
    -k(2R- r )\vu{r}  - \gamma (\Dot{\vb{r}}  \cdot \vu{r} ) \vu{r} , & r  \leq 2R \\
    0, & r  > 2R
    \end{cases} \label{eq:spring}
\end{equation}
where $k$ is the spring constant, $\gamma$ is a  damping constant, and $R$ is the radius of the spheres. 

Using this damped spring approximation allows the contact force to become attractive, which is unphysical \cite{Muller2011}. However, in driven collisions with low damping, this issue can be compensated by the choice of the damping parameter \cite{Bartz2022b}. Further, we are more concerned with the final state when the collisions have ended, which is unaffected by the details of the contact force.

% The equation of motion when the magnets are in contact is \todo{this doesn't include full force}
% \begin{equation}
%     M\Ddot{\vb{r}} =-k(2R-r )\vu{r} -\gamma(\Dot{\vb{r}} \cdot\vu{r} )\vu{r} +\vb{F}_{dip}, \label{eq:EOM}
% \end{equation}
% where $M$ is the mass of the spheres, and $\vb{F}_{dip}$ is the dipole-dipole force (\ref{eq:FullForceDot})

\subsection{Non-dimensionalization}
To make the analysis more universal, we switch to dimensionless coordinates. 
The collision duration is the shortest time scale involved in this analysis, making it a good choice of time scale for the numerical simulation.
The characteristic time is defined in terms of the spring force
\begin{equation}
    t_c=\sqrt{\frac{m}{k}}=\omega_0^{-1}.
\end{equation}
We choose a  characteristic length scale that relates the dipole-dipole interaction and the spring force
\begin{equation}
    x_c=\left( \frac{ 3\mu_0 \mu_1 \mu_2}{4\pi  k} \right)^{1/5}.
\end{equation}

With these characteristic values, we define dimensionless coordinates
$\vb{r}_i \rightarrow {\vb{r}_i}/{x_c}$,
    $t \rightarrow {t}/{t_c}$.
Changing to these dimensionless variables, the equations of motion when the spheres are in contact become
\begin{eqnarray}
     \Ddot{\vb{r}}_1 &=&-(2\tilde{R}-r )\vu{r} -2\Gamma(\Dot{\vb{r}} \cdot\vu{r} )\vu{r} \nonumber \\
     &+& \frac{1}{r^4} \Big[ (   \bm{\hat{\mu}}_1 \cdot     \bm{\hat{\mu}}_2)\vu{r}
     -5(    \bm{\hat{\mu}}_1 \cdot \vu{r})(    \bm{\hat{\mu}}_2 \cdot \vu{r}) \vu{r}
     + (    \bm{\hat{\mu}}_1 \cdot \vu{r})    \bm{\hat{\mu}}_2 + (    \bm{\hat{\mu}}_2 \cdot \vu{r})    \bm{\hat{\mu}}_1   \Big] , \label{eq:EOMnon-dimensional}\\
     \beta \tilde{R}^2\Ddot{\phi}_1 &=& \frac{1}{ r^3}\left[(\bm{\hat{\mu}}_2\cdot\vu{r})(    \bm{\hat{\mu}}_1 \times \vu{r}) -    \frac{1}{3}\bm{\hat{\mu}}_1 \times  \bm{\hat{\mu}}_2 \right], \label{eq:torqueNon-dimensional}
\end{eqnarray}
where $\tilde{R}=R/x_c$ and $\Gamma=\gamma/(2m\omega_0)$ are the only remaining input parameters.

We estimate the characteristic values for the 5 mm-radius NdFeB magnets used in the experimental setup \cite{Lindn2018}.
We estimate the mass as 4 grams using the density of Nd \cite{magdata}, and the spring constant  
$    k= Y {A}/{L},$ where $A$ is the cross-sectional area,  $L=2R$ is the length of the object, and $Y$ is Young's modulus for Nd, $Y=39.5\times 10^9$ N/m$^2$ \cite{magdata2}. The resulting spring constant is $k=3.1 \times 10^8$ N/m.
We calculate the dipole moment $\mu=MV = 0.46$ Am$^2$, where $M=8.8\times 10^5$ is the magnetization of the Nd magnetic material \cite{Gonazlez2016}.

We find the characteristic $t_c=3.6 \times 10^{-6}$ s and $x_c= 7.2\times 10^{-4}$ m. 
We use $\tilde{R}=6.94$ to approximate the physical characteristics of the magnets used in the experiment. The results presented here use an initial velocity of $5.78 \times 10^{-3}$ in dimensionless units, which corresponds to 1.16 m/s using these values of $x_c$ and $t_c$.

\subsection{Numerical techniques}

We solve the equations of motion using the Euler-Cromer method, which conserves energy in oscillatory motion \cite{Cromer1981}.
This method is chosen instead of more-accurate methods, as its derivation is more easily understood by introductory students. 
The translational (\ref{eq:EOMnon-dimensional}) and rotational (\ref{eq:torqueNon-dimensional}) equations of motion are solved in parallel. The torque (\ref{eq:torqueB}) is already accounted for by the force equation, and does not need to be solved separately.

The simulations are performed using GlowScript, \cite{glowscript} a free online tool to run VPython in the browser. VPython is designed for animation, as seen in Figure \ref{fig:Sphere_Animation}, which aids in student understanding. Three-dimensional vector operations are also built in, which simplifies the programming techniques that students must learn.

Students with more computational experience could improve the performance of this simulation by using more-advanced numerical techniques. 
Accuracy can be improved by using  higher-order differential equation solvers, and simulations can be sped up by removing loops in favor of the built-in vectorization methods in NumPy or MATLAB.

% It was necessary to define certain parameters for the sphere such as the size of the spheres, their initial starting positions, as well as their starting velocity. Both of the spheres contained equal conditions to one another, so as certain values of the program were changed to ball one, the same effect would take place on ball two.

\begin{figure}
    \centering
    \includegraphics[width=0.8\textwidth]{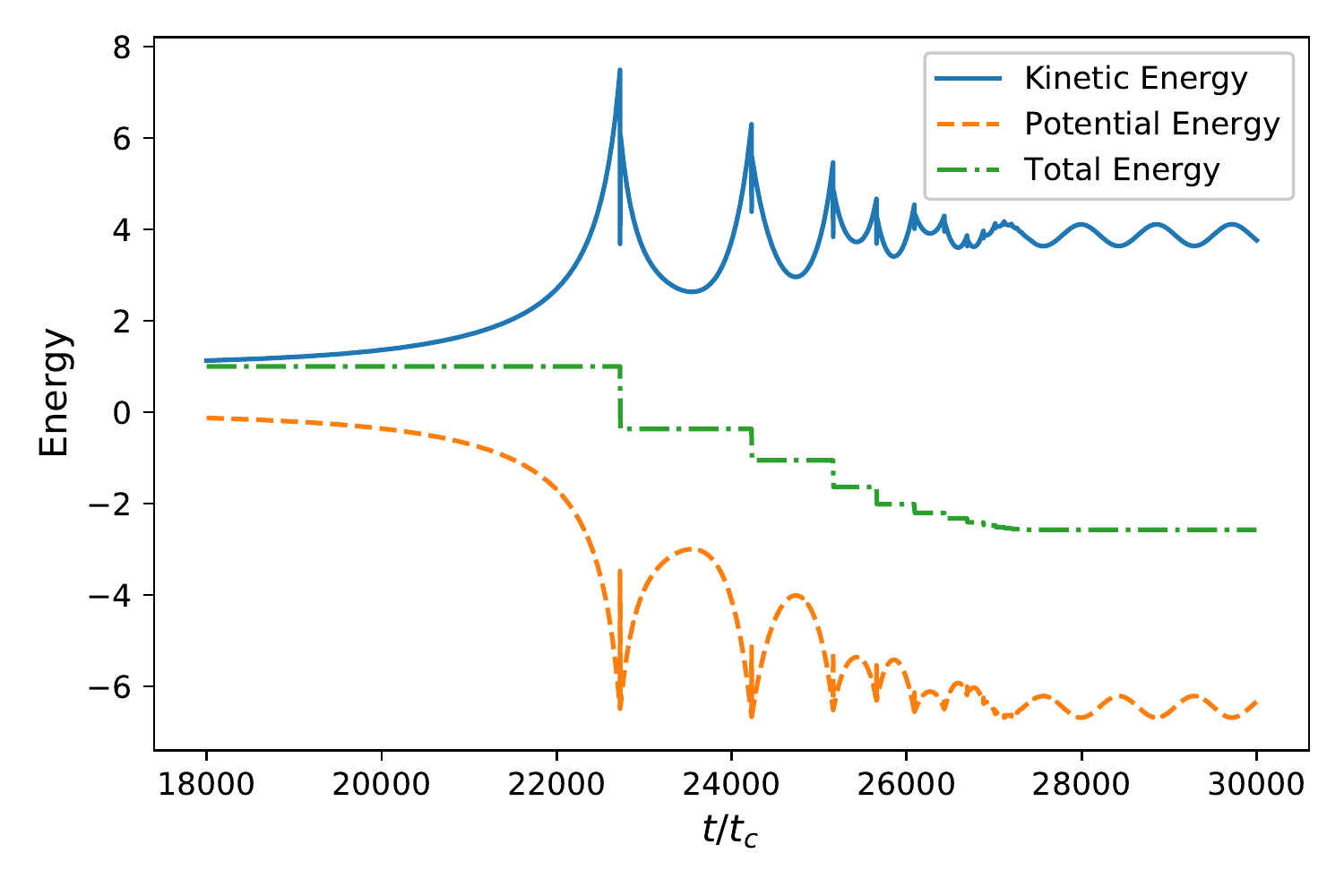}
    \caption{The time evolution of the energy for a typical solution is shown. Here, $b/R=4.5,$ and the initial velocity of each sphere is $4.3 \times 10^{-3}$ in dimensionless units. The energies are scaled by the initial energy of the system, and the potential energy includes both the magnetic (\ref{eq:PE}) and elastic energies. }
    \label{fig:KEPEfull}
\end{figure}

\section{Results}
% In comparison to the experimental inspiration, we check to ensure that the numerical analysis conserves angular momentum. We also perform the energy analysis suggested in that reference. \cite{Lindn2018}
%The spheres have identical radii $\tilde{R} = 6$, to correspond to the 5 mm magnets used in the experiment.

In each simulation,  the magnets start far away from each other, such that $U_i/K_i \ll 1$, and we initially align both magnetic moments with $\vu{r} $, as shown in Figure \ref{fig:initialConditions}. The initial kinetic energy is fixed while the impact parameter is varied. We keep $\Gamma =0.05$ fixed, although the final state does not depend on the details of the damping. 

The time evolution of the energy of the system is shown in Figure \ref{fig:KEPEfull}. 
Energy is dissipated from the system in each collision. 
Although the total energy is initially positive, the spheres will stick together only if the damping is sufficient to make the total energy negative after the first collision.

The total energy of the system reaches a constant value when the collisions cease, and the spheres remain in contact while orbiting.
If the magnets truly formed a rigid barbell shape, then the orbital and spin angular velocities would be identical, that is, $\Dot{\phi}_i=\Dot{\theta}$, and the potential energy (\ref{eq:PEangles}) would be constant. 
In particular, the magnetic moments would align with $\vu{r}$, and $\phi_i(t)=\theta_i(t)$. In this case, the potential energy reduces to
\begin{equation}
    U_{12}=-\frac{\mu_0\mu_1\mu_2}{16\pi R^3}, 
\end{equation}
with the kinetic energy also constant. 

\begin{figure}
    \centering
    \includegraphics[width=0.8\textwidth]{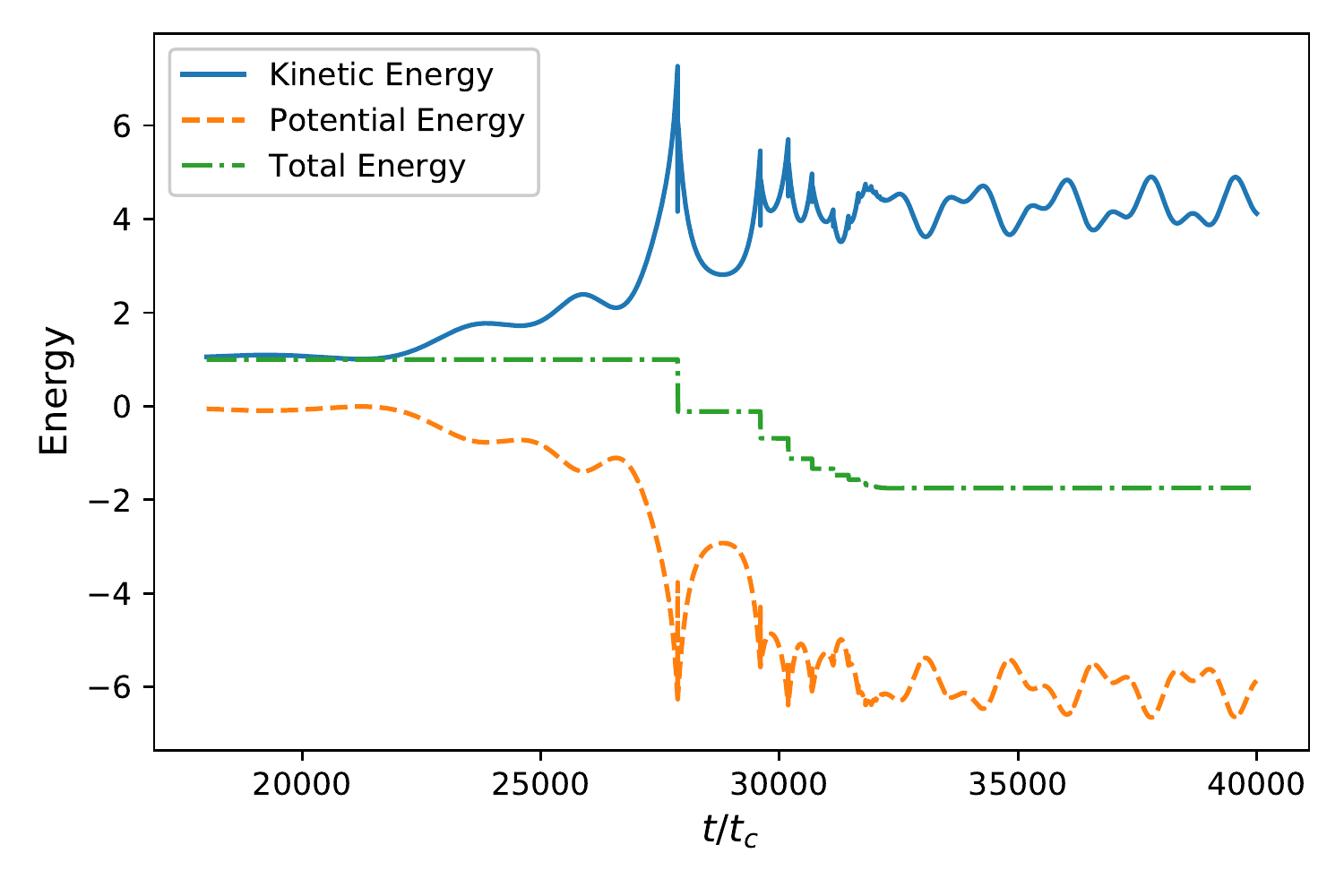}
    \caption{The time evolution of energy is shown for conditions identical to that of Figure \ref{fig:KEPEfull}, except the magnetic moments start out anti-aligned. The quasi-periodic nature of the oscillations in kinetic and potential energy is evident.}
    \label{fig:quasiperiodic}
\end{figure}

However, the alignments of the magnetic moments and $\vu{r}$ oscillate around this equilibrium state, 
so the kinetic and potential energies also fluctuate, as seen in Figure \ref{fig:KEPEfull}. 
This motion is described in terms of coupled orbital and sliding modes \cite{Haugen2020}, which produce quasi-periodic oscillations \cite{Edwards2017b}.
The quasi-periodic nature is more easily seen in Figure \ref{fig:quasiperiodic}, in which the magnetic moments are anti-aligned in the initial conditions. The final total energy also differs in this case. Further examination of these initial conditions is suggested as a future student project.

The oscillations in the final state are small and rapid compared to the barbell shape's rotational period. 
We mimic the experimental method for measuring the
 final rotational frequency  by averaging the periods of the first 5 revolutions of the barbell after the spheres have stopped bouncing. We also calculate an uncertainty from the standard deviation of these periods.
 
\begin{figure}
    \centering
    \includegraphics[width=0.8\textwidth]{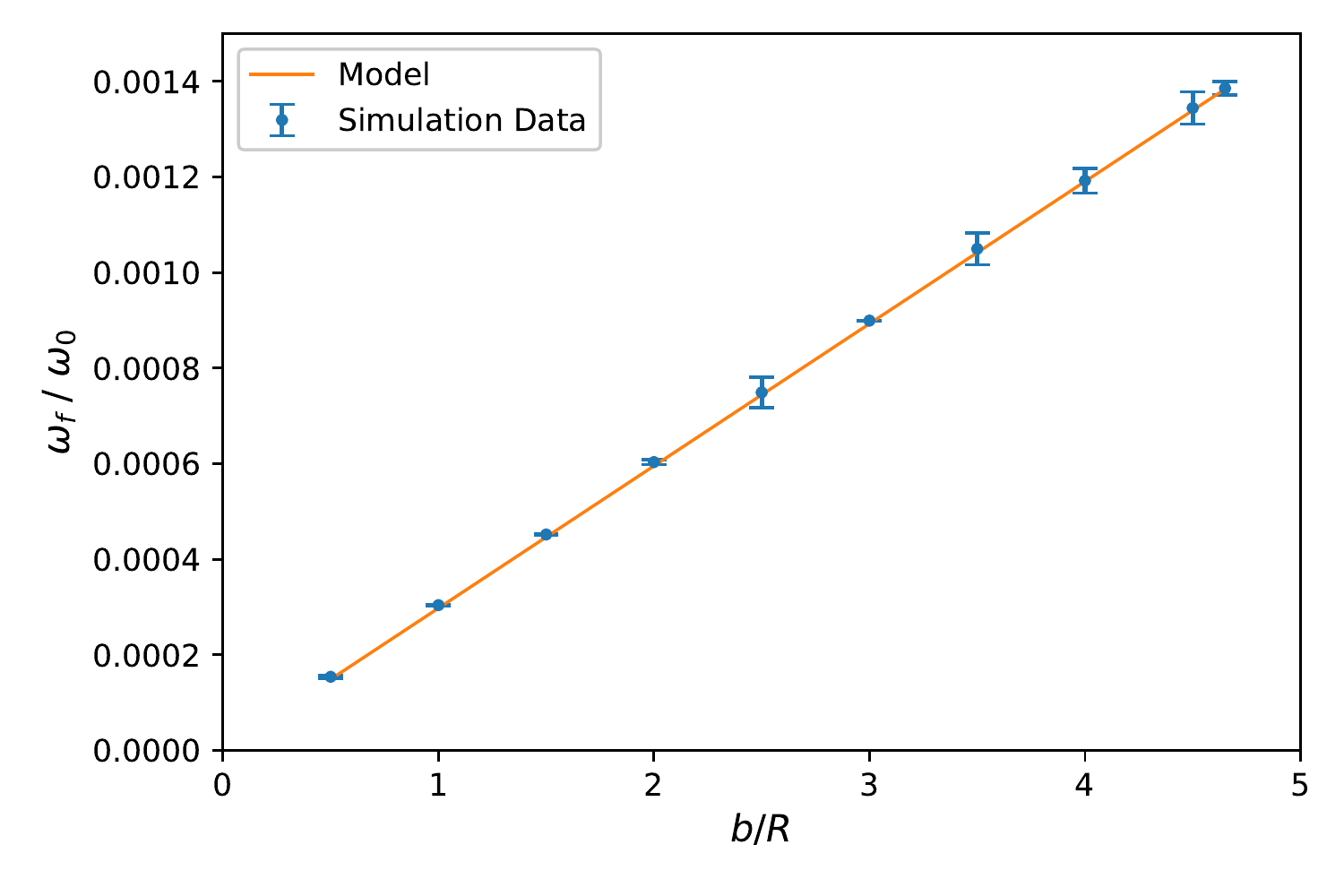}
    \caption{The final angular velocity is plotted vs the impact parameter $b/R$. The fit to this line using  (\ref{eq:omegaf}) gives $\beta=2.80$, in agreement with prediction and the experimental measurements in ref \cite{Lindn2018}. The model fit is shown. The maximum value of $b/R =4.65,$ as the magnets do not collide for larger values at the initial velocity used in these simulations. }
    \label{fig:omega}
\end{figure}

\begin{figure}
    \centering
    \includegraphics[width=0.8\textwidth]{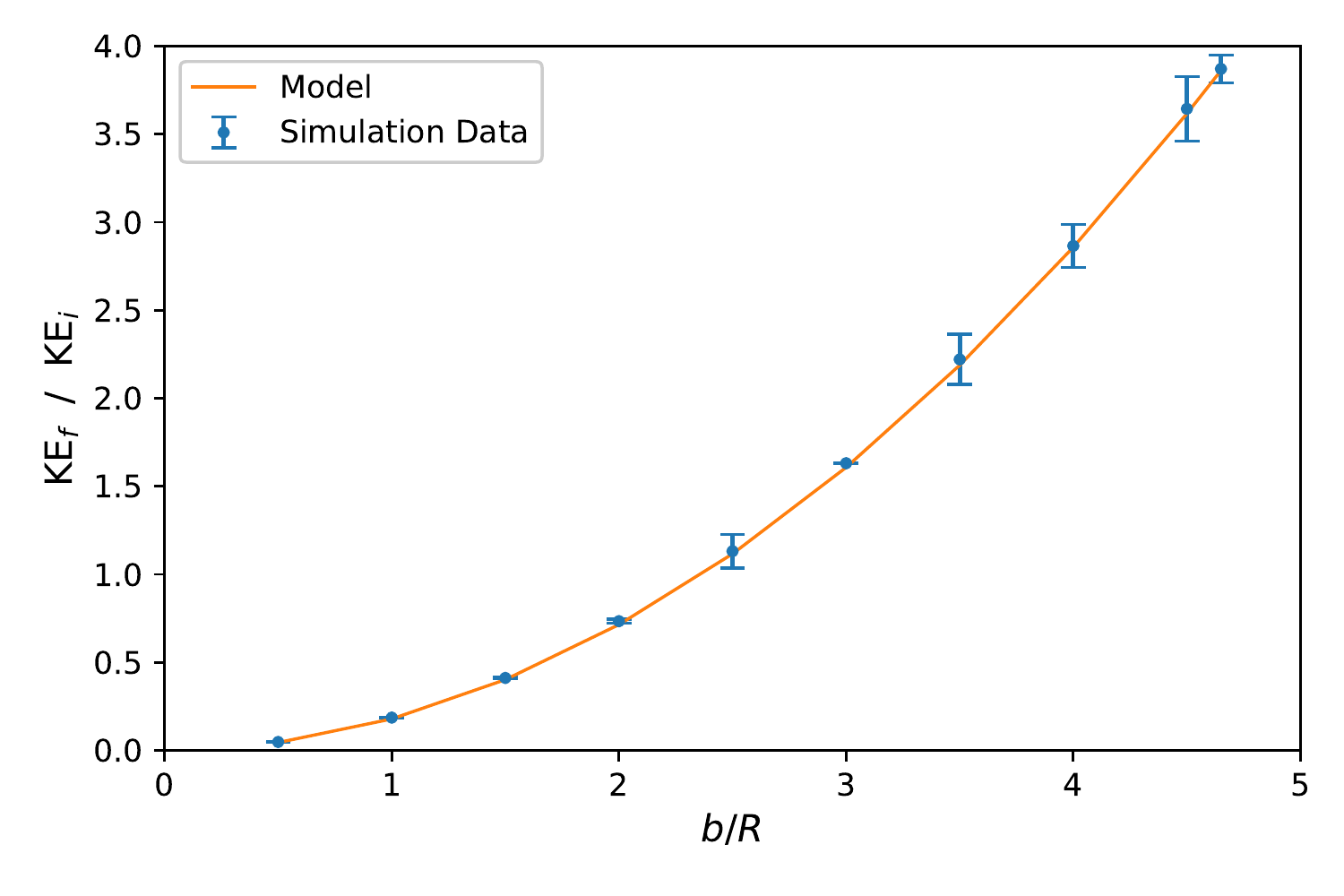}
    \caption{The ratio of final to initial kinetic energy depends on the impact parameter $b/R$. The quadratic fit matches the prediction (\ref{eq:KEratio}). In several cases, the final kinetic energy exceeds the initial, due to the change in magnetic potential energy. The model fit is shown.}
    \label{fig:KEratio}
\end{figure}

The rotational frequency is plotted as a function of impact parameter $b/ R$ in Figure \ref{fig:omega}. Fitting this line, we calculate $\beta=2.80$ from \eqref{eq:omegaf}, matching the prediction for the barbell shape.
We calculate the average final kinetic energy from this angular velocity. The plot in Figure \ref{fig:KEratio} matches the prediction \ref{eq:KEratio}. The final kinetic energy exceeds the initial when $b/R>\sqrt{2\beta}$, highlighting the work done by the magnetic field on the magnets.

When the magnets are in their final state, they are somewhat compressed, so their separation $r  < 2R$.
However, for the parameters studied here, 
\begin{equation}
    \frac{2R-r }{2R} \sim 10^{-5},
\end{equation}
so calculations based on the rigid barbell shape will not be noticeably affected by this compression.

\subsection{Simplified dipole-dipole force}\label{simple}
% For pedagogical purposes, we consider a special case where both dipole moments remain aligned with $\vb{r} $, and the dipole-dipole force simplifies to 
The lowest energy configuration occurs when the dipoles align in the direction of $\vu{r} $. In this case, the force equation becomes
\begin{equation}
\vb{F}_1=-\frac{3\mu_0m_1m_2}{2\pi r ^4}  \vu{r}  . \label{eq:simple}
\end{equation}
With this approximation, all forces in (\ref{eq:EOMnon-dimensional}) are central, and no torques are present. Thus, angular momentum is trivially conserved, and the numerical program reflects this within the precision allowed by floating-point calculations.

However, the final angular velocities found in the simulation  match (\ref{eq:omegaf}) with $\beta=2$, corresponding to treating the spheres as point masses. In the absence of non-central forces, the two magnets do not behave as a rigid barbell.
This value of $\beta$, obtained from a linear fit to the simulation data, allows us to reject the power law approximation for the dipole-dipole force,  illustrating the importance of the non-central terms in describing the motion.
% \begin{figure}
%     \centering
%     \includegraphics[width=0.8\textwidth]{Kf:Ki vs impact squared.png}
%     \caption{The ratio of final to initial kinetic energy depends on the impact parameter and the numerical factor in the moment of inertia, $\alpha$. The fit to this plot gives $\alpha=2.00$, which corresponds to point particles, not solid spheres. }
%     \label{fig:my_label}
% \end{figure}

\section{Conclusions}
In this paper, we numerically model the non-central collision of two spherical magnets, finding results in agreement with experimental observation and theoretical prediction. 
We find that the dipole-dipole force cannot be approximated as a power law, but the full force expression including non-central terms does accurately model the final state of the magnets. 
The details of the contact force are less important, as long as repulsion and dissipation are present.

This project arose from a sophomore-level course on techniques of approximation, numerical calculation, and data analysis. 
The student (JS) selected an experimental paper upon which to build a model. 
An iterative process was emphasized, beginning with the simplest approximation, and adding more details when the simple model failed to accurately reflect known results. 
Several techniques were learned in a ``just-in-time" manner, motivated by the needs of the project, as guided by the instructor (SB). 
Of particular benefit was the technique of non-dimensionalization, which helped tame the seemingly overwhelming number of independent parameters in the problem.
Another important learning outcome was evaluating the model's validity  by determining $\beta$ from the best fit to simulation data.
The use of VPython was also noted as beneficial to a novice programmer, as the animations were helpful in seeing progress and in debugging unphysical results. 

% More general model-building strategies were emphasized throughout the project. We began with the simple power-law approximation 

This work suggests several possibilities for future student projects. 
At large velocities and impact parameters, the magnets pass each other without making contact.
The initial orientation of the magnetic moments can also affect whether the magnets collide.
We restricted ourselves to initial conditions that allow the spheres to make contact and stick together. 
The initial conditions that result in collisions can be explored theoretically and computationally.
The magnets will end up stuck together if sufficient energy is dissipated in the initial collision, which depends upon the damping parameter $\Gamma$ and on the initial conditions. 

Extending from magnets, a similar analysis can be extended to any potential with long-range attraction and short-range repulsion. 
If these potentials are central, an additional contact torque, such as friction,  \cite{Schwager2008} must be included to produce the expected angular frequencies for a rigid final state. 
Such a model could be analogized to a rudimentary simulation of clumping in granular flow or molecular dynamics.
Friction could also be included in the dipole-dipole simulation. Students could check that the oscillations around the final barbell state would be suppressed.

\bibliography{references.bib}
\bibliographystyle{utphys}

\end{document}